\documentclass[graybox]{svmult}

\usepackage{mathptmx}       
\usepackage{helvet}         
\usepackage{courier}        
\usepackage{type1cm}        
\usepackage{makeidx}         
\usepackage{graphicx}        
\usepackage{multicol}        
\usepackage[bottom]{footmisc}
\usepackage{amsfonts}

\newcommand{\beq}{\begin{equation}}
\newcommand{\eeq}{\end{equation}}
\newcommand{\beqa}{\begin{eqnarray}}
\newcommand{\eeqa}{\end{eqnarray}}
\newcommand{\nn}{\nonumber \\}

\def \disk {\mathrm{disk}}
\def \e {\mathrm{e}}

\def \la {\langle}
\def \ra {\rangle}
\def \s {\sigma}
\def \t {\tau}

\def \B {{\mathcal B}}

\def \H {{\mathcal H}}

\def \Z {{\mathbb Z}}
\def \ch {\mathrm{ch}}

\def \el {\mathrm{el}}

\def \z {\zeta}
\def \L {\Lambda}
\def \D {\Delta}
\def \I {{\mathbb I}}

\def \Im {\mathrm{Im} \, }

\def \mod {\ \mathrm{mod} \ }
\def \H {{\mathcal H}}
\def \uu {{\widehat{u(1)}}}
\def \P {{\mathcal P}}
\begin{document}

\title*{Thermoelectric characteristics  of ${\mathbb Z}_k$ parafermion Coulomb islands}
\author{Lachezar S. Georgiev}
\institute{Lachezar S. Georgiev \at Institute for Nuclear Research and Nuclear Energy, Bulgarian Academy of Sciences, 
72 Tsarigradsko Chaussee, 1784 Sofia, Bulgaria, \email{lgeorg@inrne.bas.bg}}
\maketitle
\abstract{
Using the explicit rational conformal field theory partition functions for the ${\mathbb Z}_k$ parafermion 
quantum Hall states on a disk we compute numerically the thermoelectric power factor for Coulomb-blockaded islands 
at finite temperature.
We demonstrate that the power factor is rather sensitive to the neutral degrees of freedom and 
could eventually be used to distinguish experimentally between different quantum Hall states 
having identical electric properties. This might help us to confirm whether non-Abelian 
quasiparticles, such as the Fibonacci anyons, are indeed present in the experimentally observed
quantum Hall states.
}
\section{Introduction: non-Abelian anyons and Topological Quantum Computation}
\label{sec:NA}
We shall start this section with the question of what non-Abelian statistics is.
It is well known that when we exchange indistinguishable particles the quantum state acquires a phase $\e^{i\pi (\theta/\pi)}$ 
which is proportional to the statistical angle $\theta/\pi$. In three-dimensional coordinate space this phase can be either 
0 when the particles are bosons,or 1 when the particles are fermions. In two-dimensional space, however, 
this restriction is not valid and the particles can have any statistical angle between 0 and 1, that's why they are called anyons.
For example, the Laughlin anyons corresponding to the fractional quantum Hall (FQH) state with filling factor $1/3$ have  $\theta_{L}/\pi=1/3$.
In addition, while the $n$-particle quantum  states in three dimensions are constructed as  representations of the symmetric group 
$\mathcal{S}_n$, which are symmetric for bosons and antisymmetric for fermions, in two dimensions the $n$-particle states are 
build up as representations of the braid group $\B_n$. The \textit{non-Abelian anyons} are such particles in two dimensional space
whose $n$-particle states belong to representations of $\B_n$ whose dimension is bigger than 1.
In terms of $n$-particle states  this means that the non-Abelian anyons'  wave functions belong to degenerate multiplets and that  
the statistical angle $\theta$ may be a non-trivial matrix, in which case the statistical phase $\e^{i\theta}$  would be non-Abelian.

The anyonic states of matter are labeled by fusion paths \cite{clifford} which are defined as concatenation of fusion channels  and can be 
displayed in Bratteli diagrams. The fusion channels are denoted by the index `$c$' in the fusion process of two particles of type $a$ and $b$
which is denoted symbolically as $\Psi_a \times   \Psi_b =   \sum\limits_{c=1}^{g}   N_{ab}{}^c   \Psi_c $, where the fusion 
coefficients  $(N_{ab})^c$ are integers which are symmetric and associative \cite{CFT-book}. 
Put another way, a collection of particles $\{ \Psi_a \}$ are non-Abelian anyons if
$N_{ab}{}^c \neq 0$ for more than one $c$. As an example we consider the Ising anyons $\Psi_I(z)=\s(z)\ :\e^{i\frac{1}{2\sqrt{2}}\phi(z)}:$ 
realized in a conformal field theory (CFT) with $\uu \times \mathrm{Ising}$ symmetry, where $\phi(z)$ is a normalized $\uu$ 
boson and $\s$ is the chiral spin field in the Ising CFT model,
\[
\s \times \s = \I +\psi .
\]
Besides the fact that non-Abelian statistics is a new fundamental concept in particle physics it is also important
for the so-called topological quantum computation (TQC) \cite{nielsen-chuang, sarma-RMP}. In this context quantum information is encoded 
into the fusion channels 
\beqa
|0\ra=(\s,\s)_{\I}\quad\ &\longleftrightarrow& \quad \s \times \s  \to \I \nn
|1\ra=(\s,\s)_{\psi}\quad &\longleftrightarrow& \quad \s \times \s  \to \psi \nonumber ,
\eeqa
which is a topological quantity--it is is independent of the fusion process details depending only on the topology 
of the coordinate space. Fusion channel is also  independent of the anyon separation and is preserved when the two particles are 
separated--if we fuse two particles and then split them again,  their fusion channel does not change. 

The basic idea of TQC is that quantum information can be encoded into the fusion channels and the quantum gates can be implemented 
by braiding non-Abelian anyons. As an illustration we can consider 8 Ising anyons, which in the quantum information language can be used to 
encode 3 topological qubits, and transport adiabatically anyon number $7$  along a complete loop around anyon number $6$.
Then the 8-anyons states are multiplied by a statistical phase $\left(B^{(8,+)}_6\right)^2= X_3$ which implements the  
NOT gate $X_3= \I_2\otimes\I_2 \otimes X$ on the third qubit \cite{ultimate,nielsen-chuang}.

Another promising example  of non-Abelian anyons are the Fibonacci anyons \cite{fibonaci} realized in the diagonal coset of the 
$\Z_3$ parafermion FQH states \cite{NPB2001,NPB2015-2} (or, in the three-state Pots model) as the the parafermion primary 
field $\epsilon$ corresponding to the nontrivial orbit of the simple-current's action $\I =\{ \L_0+\L_0,\L_1+\L_1,\L_2+\L_2 \}$ 
$\epsilon =\{ \L_0+\L_1,\L_1+\L_2,\L_0+\L_2 \}$  with fusion rules
\[
\I \times \I = \I , \quad \I \times \epsilon = \epsilon, \quad \epsilon \times \epsilon = \I +\epsilon .
\]
The information encoding for Fibonacci anyons is again in the fusion channels, denoted by the field $\I$ and $\epsilon$ of the resulting 
fusion, however, this time  for triples of anyons \cite{fibonaci}
\beqa
|0\ra &=&((\epsilon,\epsilon)_{\I} , \epsilon)_{\epsilon}  \nn
|1\ra &=&((\epsilon,\epsilon)_{\epsilon}, \epsilon)_{\epsilon}  \nonumber ,
\eeqa
and the third state $((\epsilon,\epsilon)_{\epsilon}, \epsilon)_{\I}$, having a trivial quantum dimension,
decouples from the previous two and is called non-computational  \cite{fibonaci}.

Given that non-Abelian statistics is a new concept a natural question arises: how can it be discovered?
In the rest of this paper we will discuss, how non-Abelian statistics might be observed in
Coulomb-blockade conductance spectrometry and by measuring certain thermoelectric characteristics of Coulomb-blockaded 
islands.
\section{Coulomb island spectroscopy}
\label{sec:CB}
Let us consider a Coulomb-blockaded island, which can be realized as a quantum dot with drain, source and a side gate
which is equivalent to single-electron transistor, like in Ref.~\cite{LT10}.
This setup is an almost closed quantum system, which still has  discrete energy levels and is like a
large artificial atom but is highly tunable by Aharonov--Bohm flux and side-gate voltage.
\subsection{Coulomb island's conductance--CFT approach}
In this section we are going to use the chiral Grand canonical partition function for a disk fractional quantum Hall sample
to calculate its thermoelectric properties. In such samples the bulk is inert due to the nonzero mobility gap while the edge is 
mobile and can be described by a rational unitary CFT  \cite{fro-stu-thi,cz,CFT-book}. The Grand partition function is
\beq
Z_{\disk}(\t,\z) = \mathrm{tr}_{ \H_{\mathrm{edge}}} \ \e^{-\beta (H-\mu N)} 
= \mathrm{tr}_{ \H_{\mathrm{edge}}} \ \e^{2\pi i \t (L_0 -c/24)} e^{2\pi i \z J_0}, 
\eeq
 where $H=\hbar\frac{2\pi v_F}{L} \left(L_0-\frac{c}{24}\right)$ is the edge Hamiltonian expressed in terms of the  zero mode $L_0$
of the Virasoro  stress-energy tensor (with a central charge $c$),  $N=-\sqrt{\nu_H} J_0$ is the particle number on the edge expressed
 in terms of the zero $J_0$ mode of the $\uu$ current and $\nu_H$ is the  FQH filling factor. 
The trace is taken over the edge-states' Hilbert space $\H_{\mathrm{edge}}$ 
which depends on the number of quasiparticles localized in the bulk.
The temperature $T$ and chemical potential $\mu$ are related to the modular parameters \cite{CFT-book} $\t$ and $\z$ by
 \[
\t=i\pi\frac{T_0}{T}, \quad  T_0=\frac{\hbar v_F }{\pi k_B L}, \quad \z= i\frac{\mu}{2\pi  k_B T} ,
\]
where $v_F$ is the Fermi velocity at the edge and $L$ is the circumference of the edge.
The CFT disk partition function  in presence of AB flux $\phi=eB.A/h$, threading the disk, is modified by simply shifting the 
chemical potential  \cite{NPB-PF_k}
\[
\z \to \z +\phi \t, \quad Z_{\disk}^{\phi}(\t,\z ) = Z_{\disk}(\t,\z +\phi\t).
\]
It is interesting to note that the side-gate voltage $V_g$ affects the quantum dot (QD) in the same way as the AB flux \cite{NPB2015}, 
through the externally induced electric charge \cite{matveev-LNP}  on the QD (continuous)  
 $-C_g V_g/e\equiv \nu_H \phi =Q_{\mathrm{ext}} $,
where $C_g$ is the capacitance of the gate.

The thermodynamic Grand potential on the edge is defined as usual as $\Omega(T,\mu)=-k_B T \ln Z_{\disk}(\t,\z)$  and
the electron number can be computed as  \cite{thermal} 
\beq\label{N-av}
\la N_\el (\phi)\ra_{\beta,\mu_N} 
 =\nu_H\left(\phi+ \frac{\mu_N}{\Delta \epsilon} \right) +\frac{1}{2\pi^2} \left(\frac{T}{T_0}\right) \frac{\partial }{\partial \phi}
 \ln Z_\phi(T,\mu_N) 
\eeq
Similarly the edge conductance in the linear-response regime can be computed by  \cite{thermal} 
\beq\label{G}
G (\phi)=\frac{e^2}{h}
\left( \nu_H +\frac{1}{2\pi^2} \left(\frac{T}{T_0} \right)\frac{\partial^2 }{\partial \phi^2}  \ln Z_{\phi}(T,0)\right) .
\eeq
There are certain difficulties in measuring QD conductance and distinguishing FQH states: 
experiments are performed in extreme conditions (high $B$, very low $T$) 
with expensive samples. Moreover, there are many doppelgangers \cite{nayak-doppel-CB}, i.e., distinct states with 
the same conductance patterns 
with differences in the neutral sector where $G$ not sensitive. 
Under these conditions the sequential tunneling of electrons one-by-one is dominating the cotunneling, which is a 
higher-order process associated with almost simultaneous virtual tunneling of pairs of 
electrons \cite{matveev-LNP}, that will not be considered here.
\section{Thermopower: a finer spectroscopic tool  }   
The thermopower, or the Seebeck coefficient, is defined \cite{matveev-LNP,NPB2015} as the potential difference $V$ between the leads of the SET 
when $\Delta T=T_R -T_L\ll T_L$, under the condition that $I=0$. Usually thermopower is expressed as $S=G_T/G$, where $G$ and 
$G_T$ are electric and thermal conductances, respectively. However, for the SET configuration
 $G_T\to 0$ and $G\to 0 $ in the  Coulomb blockade valleys, so that it is more appropriate to use another expression \cite{matveev-LNP}
\[
S \equiv  \left. -\lim_{\Delta T \to 0} \frac{V}{\Delta T}\right|_{I=0}=
-\frac{\la \varepsilon \ra}{eT}, \quad
\]
where  $ \la \varepsilon \ra$ is the average energy of the  tunneling electrons.
In the CFT approach using the Grand partition function for the FQH edge of the QD we can express the average tunneling energy
as the difference between the energy of the QD with $N+1$ electrons and that of the QD with $N$ electrons
\[
\la \varepsilon \ra^{\phi}_{\beta,\mu_N} = \frac{E^{\beta,\mu_{N+1}}_{\mathrm{QD}}(\phi)-  E^{\beta,\mu_{N}}_{\mathrm{QD}}(\phi)  }
{ \la N(\phi)\ra_{\beta,\mu_{N+1} }  -  \la N(\phi)\ra_{\beta,\mu_{N}}}
\]
The total QD energy  (in the Grand canonical ensemble) can be written as
\[
E^{\beta,\mu_{N}}_{\mathrm{QD}}(\phi) =\sum_{i=1}^{N_0} E_i+\la H_{\mathrm{CFT}}(\phi)\ra_{\beta,\mu_{N}} ,
\]
where $E_i$, $i=1, \ldots , N_0$ are the occupied single-electron states in the bulk of the QD,  
and  $\la \cdots \ra_{\beta,\mu}$ is the Grand canonical average of $H_{\mathrm{CFT}}$ on the edge at inverse temperature
$\beta=(k_B T)^{-1}$ and chemical potential $\mu$.
The chemical potentials $\mu_{N}$ and $\mu_{N+1}$ of the QD with $N$ and $N+1$ electrons can be chosen as \cite{NPB2015}
\[
\mu_N=-\frac{1}{2}\Delta\epsilon, \quad \mu_{N+1}=\frac{1}{2}\Delta\epsilon,   \quad \Delta\epsilon=\hbar\frac{2\pi v_F}{L} .
\]
Another important observable is the thermoelectric power factor\cite{NPB2015} $\P_T$ which is defined as the
electric power $P$ generated by the temperature difference $\Delta T$
\beq \label{P_T}
P=V^2 / R = \P_T (\Delta T)^2, \quad \P_T= S^2 G,
\eeq
where $R=1/G$ is the electric resistance of the CB island.
The average tunneling energy can be expressed in terms of the CFT averages of the Hamiltonian and particle number as follows
 \cite{NPB2015}
\beq \label{eps}
\la \varepsilon \ra^{\phi}_{\beta,\mu_N} = 
\frac{ \la H_{\mathrm{CFT}}(\phi)\ra_{\beta,\mu_{N+1} }  -  \la H_{\mathrm{CFT}}(\phi)\ra_{\beta,\mu_{N} }}
{ \la N(\phi)\ra_{\beta,\mu_{N+1} }  -  \la N(\phi)\ra_{\beta,\mu_{N}}}
 \eeq
The electron number average can be compured from Eq.~(\ref{N-av}) and the edge energy average can be obtained from the 
Grand potential $\Omega_{\phi}(T,\mu_N)=-k_B T \ln Z_{\phi}(T,\mu)$  in presence of AB flux $\phi$ as
\beq \label{H}
\la H_{\mathrm{CFT}}(\phi)\ra_{\beta,\mu_N} =\Omega_{\phi}(T,\mu_N)- T \frac{\partial \Omega_{\phi}(T,\mu_N)}{\partial T}  
 - \mu_N\frac{\partial  \Omega_{\phi}(T,\mu_N)}{\partial \mu} .
\eeq	
\section{ $\Z_k$ parafermion quantum Hall islands  }
The CFT for the $\Z_k$ parafermion quantum Hall islands (or QDs) contains an electric charge part $\uu$
and a neutral part which is realized as a parafermion diagonal coset \cite{NPB2001}
\[
\left(\uu \otimes \frac{\widehat{su(k)}_1\oplus\widehat{su(k)}_1 }{\widehat{su(k)}_2 }\right)^{\Z_k}
 \]
The total disk partition function for the $\Z_k$ parafermion quantum Hall islands \cite{NPB2001} is labeled by two 
integers $l \mod k+2$ and  $\rho\mod k$ satisfying $l-\rho \leq \rho \mod k$ and can be written as follows
 \beq \label{Z}
\chi_{l,\rho} (\t,\z) = \sum_{s=0}^{k-1} K_{l+s(k+2)}(\t,k\z;k(k+2)) \ch(\L_{l-\rho+s}+\L_{\rho+s})(\t),
\eeq
where $K_{l}(\t,k\z;k(k+2))$ are the chiral partition functions of the $\uu$ part while $\ch(\L_{\mu}+\L_{\rho})(\t)$
are the characters of the neutral part of the CFT.
The $\uu$ part corresponds to  Luttinger liquid partition function    (with compactification radius $R_c=1/m$)
\[
 K_{l}(\t,\z; m) = \frac{\mathrm{CZ}}{\eta(\t)} \sum_{n=-\infty}^{\infty} q^{\frac{m}{2}\left(n+\frac{l}{m}\right)^2} 
\e^{2\pi i \z \left(n+\frac{l}{m}\right)} 
\] 
where the modular parameter is related to the  temperature 
\[
q=\e^{-\beta\D\varepsilon}=\e^{2\pi i \t}, \quad \D \varepsilon= \hbar\frac{2\pi v_F}{L}
\]
and the Dedekind function and Cappelli--Zemba factors \cite{cz,CFT-book} are given by
\[
 \eta(\t)=q^{1/24}\prod_{n=1}^\infty (1-q^n), \quad \mathrm{CZ}=\e^{-\pi\nu_H\frac{(\Im \z)^2}{\Im\t}}
\]
The neutral partition function   are labeled by a level-2 weight $\L_{\mu}+\L_{\rho}$ with the condition $0 \leq \mu \leq \rho \leq k-1$
and have the form \cite{NPB2001}
\[
\ch_{\s,Q}(\t)= q^{ \D(\s) - \frac{c}{24}  }
\sum\limits_{
\mathop{m_1,m_2,\ldots, m_{k-1}=0}\limits_{\sum\limits_{i=1}^{k-1}\, i
\, m_i \equiv Q \mod k } }^\infty
\frac{q^{\underline{m}. C^{-1}.
\left(\underline{m} - \L_\s \right)} }{(q)_{m_1} \cdots (q)_{m_{k-1}} },
\]
\[
(q)_n=\prod\limits_{j=1}^n (1-q^j), \quad \D(\s)=\frac{\s(k-\s)}{2k(k+2)} , \quad
c=\frac{2(k-1)}{k+2}
\]
where  $\underline{m} = (m_1,\ldots, m_{k-1})$,  \quad $0\leq \s \leq Q \leq k-1$ and $ C^{-1}$ is the inverse
 Cartan matrix for $su(k)$.
The coset weight labels are related to $\s$ and $Q$ by $\mu=Q-\s, \quad  \rho= Q$.
\begin{figure}[htb]
\centering
\includegraphics[bb=40 0 620 415,clip,width=\textwidth]{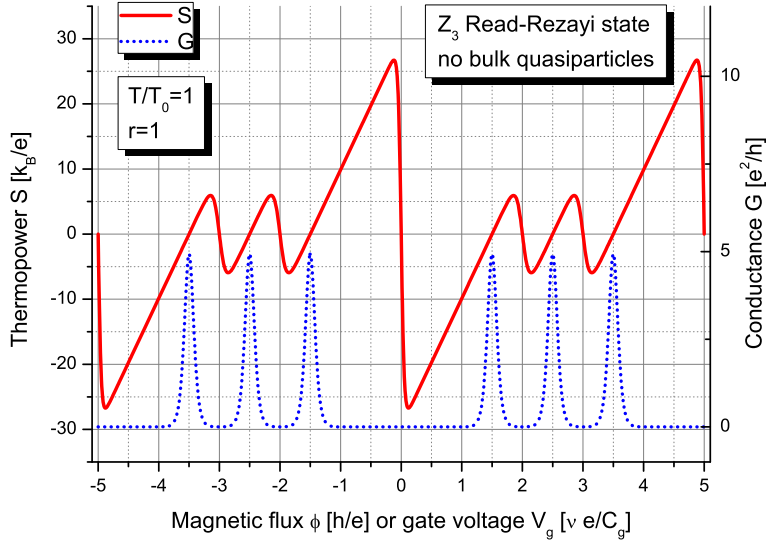}
\caption{Conductance peaks (right $Y$ scale) and thermopower (left $Y$ scale) for the $\Z_3$ 
parafermion FQH state without bulk quasiparticles. \label{fig:TP-G}}
\end{figure}
Using the explicit formulas (\ref{eps}) for the thermopower in terms of the average tunneling energy 
expressed in terms of the Grand canonical averages (\ref{N-av}) and (\ref{H}) and the partition function (\ref{Z}),
as well as Eq.~(\ref{G}) for the conductance,
we can compute the thermopower for the $\Z_3$ parafermion FQH state. We plot in Fig.~\ref{fig:TP-G}
the electric conductance and thermopower as functions of the AB flux $\phi$, or, equivalently as functions of the side-gate 
voltage $V_g$ for the $\Z_3$ parafermion FQH state without quasiparticles in the bulk, i.e., for $l=0$ and $\rho=0$.
Just like the thermopower of metallic quantum dots \cite{matveev-LNP}, we see in Fig.~\ref{fig:TP-G} that the peaks of the 
conductance precisely corresponds  to the (continuous in the limit $T\to 0$) zeros of the thermopower.

\begin{figure}[htb]
\centering
\includegraphics[bb=40 10 520 415,clip,width=\textwidth]{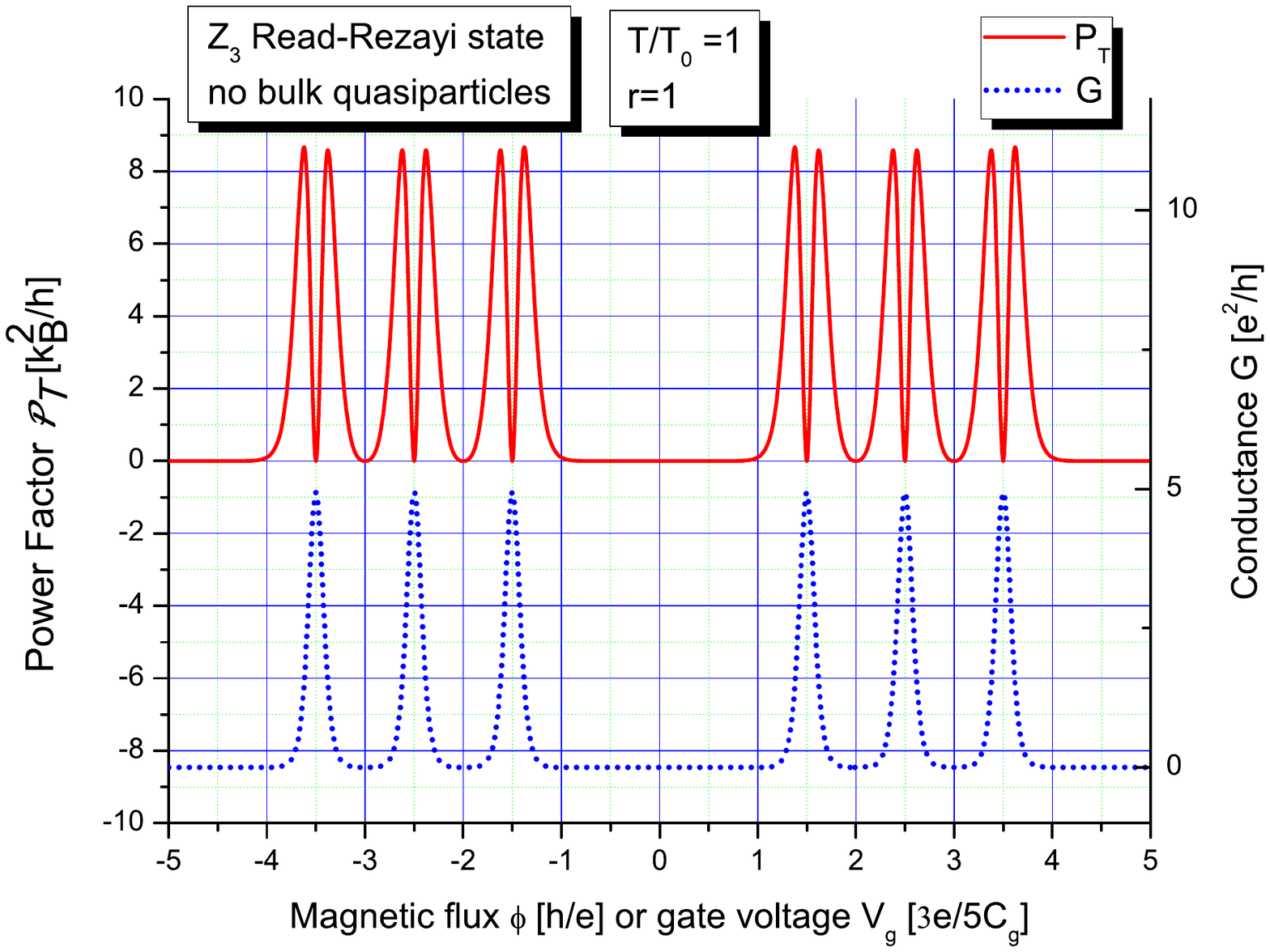}
\caption{Power factor and conductance for the $\Z_3$ parafermion FQH state without bulk 
quasiparticles. \label{fig:PF-G} }
\end{figure}

Similarly, we can compute from Eq.~(\ref{P_T}) the power factor $\P_T$ for the $\Z_3$ parafermion FQH state without 
quasiparticles in the bulk ($l=0$ and $\rho=0$), which is plotted together with the conductance in Fig.~\ref{fig:PF-G}.
\section{Conclusion and perspectives}
\label{sec:concl}
The thermoelectric characteristics of Coulomb blockaded QDs, such as the thermopower and especially the thermoelectric
power factor,  appear to be more sensitive to the neutral modes in the FQH liquid than the tunneling conductance. 
These could be used as experimental signatures to identify (non-Abelian) Fibonacci anyons\cite{fibonaci}, which are believed to 
exist in  the $\nu_H=12/5$ quantum Hall state. This experimentally observed FQH state \cite{pan-xia-08} might be a realization of the
particle--hole conjugate of the $\Z_3$ parafermion quantum Hall state in the second Landau level with  filling factor 
$\nu_H=3-k/(k+2)$ for $k=3$.

A recent experiment \cite{gurman-2-3} demonstrated that the power factor of a Coulomb blockaded quantum dot might be directly
measurable like the observable plotted in Fig.~3c there. This could allow us to estimate from the experiment the ratio 
between the Fermi velocities of the charged and neutral edge modes by comparing with the power factor profile computed 
theoretically from the CFT \cite{NPB2015-2}.
Finally, this possibility to distinguish neutral characteristics of FQH states is bringing a new hope that we could eventually 
decide  whether Fibonacci anyons are indeed realized in Nature.
\begin{acknowledgement}
I thank Andrea Cappelli, Guillermo Zemba and Ady Stern for many helpful discussions.
This work has been partially supported by the Alexander von Humboldt Foundation under the Return Fellowship and 
Equipment Subsidies Programs and by the Bulgarian Science Fund under Contract No. DFNI-E 01/2 and  DFNI-T 02/6.
\end{acknowledgement}
\bibliography{my,TQC,FQHE,Z_k,CB}

\providecommand{\href}[2]{#2}\begingroup\raggedright\begin{thebibliography}{10}

\bibitem{clifford}
A.~Ahlbrecht, L.~S. Georgiev, and R.~F. Werner, ``Implementation of {C}lifford
  gates in the {Ising}-anyon topological quantum computer,'' {\em Phys. Rev. A}
  {\bf 79} (2009) 032311, \href{http://xxx.lanl.gov/abs/arXiv:0812.2338}{{\tt
  arXiv:0812.2338}}.

\bibitem{CFT-book}
P.~{Di Francesco}, P.~Mathieu, and D.~S\'en\'echal, {\em Conformal Field
  Theory}.
\newblock Springer--Verlag, New York, 1997.

\bibitem{nielsen-chuang}
M.~Nielsen and I.~Chuang, {\em Quantum Computation and Quantum Information}.
\newblock Cambridge University Press, 2000.

\bibitem{sarma-RMP}
S.~D. Sarma, M.~Freedman, C.~N. andStiven H.~Simon, and A.~Stern,
  ``Non-{A}belian anyons and topological quantum computation,'' {\em Rev. Mod.
  Phys.} {\bf 80} (2008) 1083,
  \href{http://xxx.lanl.gov/abs/arXiv:0707.1889}{{\tt arXiv:0707.1889}}.

\bibitem{ultimate}
L.~S. Georgiev, ``Ultimate braid-group generators for exchanges of {I}sing
  anyons,'' {\em J. Phys. A: Math. Theor.} {\bf 42} (2009) 225203,
  \href{http://xxx.lanl.gov/abs/arXiv:0812.2334}{{\tt arXiv:0812.2334}}.

\bibitem{fibonaci}
N.~Bonesteel, L.~Hormozi, G.~Zikos, and S.~Simon, ``Braid topologies for
  quantum computation,'' {\em Phys. Rev. Lett.} {\bf 95} (2005) 140503.

\bibitem{NPB2001}
A.~Cappelli, L.~S. Georgiev, and I.~T. Todorov, ``Parafermion {H}all states
  from coset projections of {A}belian conformal theories,'' {\em Nucl. Phys.}
  {\bf B 599 [FS]} (2001) 499--530,
  \href{http://xxx.lanl.gov/abs/hep-th/0009229}{{\tt hep-th/0009229}}.

\bibitem{NPB2015-2}
L.~S. Georgiev, ``Thermopower and thermoelectric power factor of ${\Z}_k$
  parafermion quantum dots,'' {\em Nucl. Phys. B} {\bf 899} (2015) 289--311,
  \href{http://xxx.lanl.gov/abs/arXiv:1505.02538}{{\tt arXiv:1505.02538}}.

\bibitem{LT10}
L.~S. Georgiev, ``Thermopower in the {C}oulomb blockade regime for {L}aughlin
  quantum dots,'' in {\em Lie Theory and Its Applications in Physics,},
  V.~Dobrev, ed., Springer Proceedings in Mathematics \& Statistics 111,
  pp.~279--289.
\newblock 2014.
\newblock \href{http://xxx.lanl.gov/abs/arXiv:1406.5592}{{\tt
  arXiv:1406.5592}}.
\newblock Proceedings of the 10-th International Workshop "Lie Theory and Its
  Applications in Physics", 17-23 June 2013, Varna, Bulgaria.

\bibitem{fro-stu-thi}
J.~Fr\"{o}hlich, U.~M. Studer, and E.~Thiran, ``A classification of quantum
  {H}all fluids,'' {\em J. Stat. Phys.} {\bf 86} (1997) 821,
  \href{http://xxx.lanl.gov/abs/cond-mat/9503113}{{\tt cond-mat/9503113}}.

\bibitem{cz}
A.~Cappelli and G.~R. Zemba, ``Modular invariant partition functions in the
  quantum {H}all effect,'' {\em Nucl. Phys.} {\bf B490} (1997) 595,
  \href{http://xxx.lanl.gov/abs/hep-th/9605127}{{\tt hep-th/9605127}}.

\bibitem{NPB-PF_k}
L.~S. Georgiev, ``A universal conformal field theory approach to the chiral
  persistent currents in the mesoscopic fractional quantum {H}all states,''
  {\em Nucl. Phys.} {\bf B 707} (2005) 347--380,
  \href{http://xxx.lanl.gov/abs/hep-th/0408052}{{\tt hep-th/0408052}}.

\bibitem{NPB2015}
L.~S. Georgiev, ``{Thermoelectric properties of Coulomb-blockaded fractional
  quantum Hall islands},'' {\em Nucl. Phys. B} {\bf 894} (2015) 284--306,
  \href{http://xxx.lanl.gov/abs/arXiv:1406.6177}{{\tt arXiv:1406.6177}}.

\bibitem{matveev-LNP}
K.~Matveev, ``Thermopower in quantum dots,'' {\em Lecture Notes in Physics}
  {\bf LNP 547} (1999) 3--15.

\bibitem{thermal}
L.~S. Georgiev, ``Thermal broadening of the {C}oulomb blockade peaks in quantum
  {H}all interferometers,'' {\em EPL} {\bf 91} (2010) 41001,
  \href{http://xxx.lanl.gov/abs/arXiv:1003.4871}{{\tt arXiv:1003.4871}}.

\bibitem{nayak-doppel-CB}
P.~Bonderson, C.~Nayak, and K.~Shtengel, ``Coulomb blockade doppelgangers in
  quantum {H}all states,'' {\em Phys. Rev. B} {\bf 81} (2010) 165308,
  \href{http://xxx.lanl.gov/abs/arXiv:0909.1056}{{\tt arXiv:0909.1056}}.

\bibitem{pan-xia-08}
W.~Pan, J.~S. Xia, H.~L. Stormer, D.~C. Tsui, C.~Vicente, E.~D. Adams, N.~S.
  Sullivan, L.~N. Pfeiffer, K.~W. Baldwin, and K.~W. West, ``Experimental
  studies of the fractional quantum hall effect in the first excited landau
  level,'' {\em Phys. Rev. B} {\bf 77} (Feb, 2008) 075307.

\bibitem{gurman-2-3}
I.~Gurman, R.~Sabo, M.~Heiblum, V.~Umansky, and D.~Mahalu, ``Extracting net
  current from an upstream neutral mode in the fractional quantum {H}all
  regime,'' {\em Nature Communications} {\bf 3} (2012) 1289,
  \href{http://xxx.lanl.gov/abs/arXiv:1205.2945}{{\tt arXiv:1205.2945}}.

\end{thebibliography}\endgroup

\end{document}